# The trade-off between factor score determinacy and the preservation of inter-factor correlations

André Beauducel*, Norbert Hilger, & Tobias Kuhl

*University of Bonn, Department of Psychology*

**Abstract**

Regression factor score predictors have the maximum factor score determinacy, i.e., the maximum correlation with the corresponding factor, but they do not have the same inter-correlations as the factors. As it might be useful to compute factor score predictors that have the same inter-correlations as the factors, correlation-preserving factor score predictors have been proposed. However, correlation-preserving factor score predictors have smaller correlations with the corresponding factors (factor score determinacy) than regression factor score predictors. Thus, higher factor score determinacy goes along with bias of the inter-correlations and unbiased inter-correlations go along with lower factor score determinacy. The aim of the present study was therefore to investigate the size of the trade-off between factor score determinacy and bias of inter-correlations by means of a simulation study. It turns out that under several conditions very small gains of factor score determinacy of the regression factor score predictor go along with a large bias of inter-correlations. Instead of using the regression factor score predictor by default, it is proposed to check whether substantial bias of inter-correlations can be avoided without substantial loss of factor score determinacy by using a correlation-preserving factor score predictor. A syntax that allows to compute correlation-preserving factor score predictors from regression factor score predictors and to compare factor score determinacy and inter-correlations of the factor score predictors is given in the Appendix.

* Corresponding author, address for correspondence:
University of Bonn, Institute of Psychology, Kaiser-Karl-Ring 9, 53111 Bonn
Phone: +49228 734151, Email: beauducel@uni-bonn.de

In the context of psychological assessment, the computation of individual scores following factor analysis might be useful. The decision whether an individual should be assigned to an intervention or whether an individual should get a job does not only require a prediction model but also the scores of the individual. Accordingly, several factor score predictors have been proposed and evaluated in different contexts (Grice, 2001; DiStefano, Zhu, & Mîndrilă, 2009). A very common factor score predictor is the regression factor score predictor, which is sometimes termed best-linear factor score predictor (Thurstone, 1935; Krijnen, Wansbeek, & ten Berge, 1996). The regression factor score predictor has the maximum correlation with the factor score, which is the maximum factor score determinacy (Grice, 2001). However, the regression factor score predictor does typically not result in the same inter-correlations as the factors estimated within the factor model. However, as the factor score predictors should represent the factors of the model, it might be argued that they should have the same inter-correlations as the factors themselves. Therefore, correlation-preserving factor score predictors have been proposed by McDonald (1981). The method proposed by McDonald (1981) allows to compute correlation-preserving factor score predictors from the measured variables, but it has not the maximum factor score determinacy.

Although the theoretical differences between the regression factor score predictor and the correlation-preserving factor score predictor have been already been described and discussed (Nicewander, 2019; Grice, 2001, Krijnen, Wansbeek, & ten Berge, 1996), the trade-off between factor score determinacy and bias of the inter-correlations of factor score predictors has rarely been systematically investigated. Moreover, there are different possibilities to compute correlation-preserving factor score predictors. The correlation-preserving factor score predictor proposed by McDonald (1981) are computed from the model parameters whereas the correlation-preserving factor predictor proposed by Beauducel and Hilger (2022) is based on a transformation of a given factor score predictor. The latter can be useful in the context of Bayesian plausible values, when a software package only provides factor score predictors that are not correlation preserving and when the direct computation of the factor score predictor from model parameters might be questionable (Asparouhov & Muthén, 2010). The latter factor score predictor might also be of interest when only the factor score predictors and the model parameters are available, whereas the observed variables are unavailable. However, the correlation-preserving factor score predictor proposed by McDonald (1981) and Beauducel and Hilger (2022) have not been directly compared so that possibly different factor score determinacies are until know unknown. Thus, the current investigation is relevant for two reasons: First, because several software-packages solely compute the regression factor score predictor while the trade-off between factor score determinacy and the departure from the inter-correlations of the factors remains unknown. Second, because it is unknown under which conditions one of the two-versions of correlation-preserving factor score predictors should be preferred.

First, some definitions of the factor model, the regression factor score predictor, the correlation-preserving factor score predictor, and the transformation of the regression factor score predictor into a correlation-preserving factor score predictor is shown. Second, a simulation study is performed that allows to compare the factor score determinacy of the regression factor score predictor with the determinacy of the correlation-preserving factor score predictor. Finally, the correlation-preserving transformation is applied to the scores from an empirical data set and the consequences of the use of correlation-preserving factor score predictors are discussed.

**Definitions**

The population factor model can be written as

$$\mathbf{x} = \mathbf{\Lambda}\boldsymbol{\xi} + \boldsymbol{\Psi}\boldsymbol{\varepsilon}, \qquad (1)$$

where $\mathbf{x}$ is a vector of $p$ observed variables, $\mathbf{\Lambda}$ is a $p \times q$ matrix of factor loadings, $\boldsymbol{\xi}$ is a random vector of common factors with $E(\boldsymbol{\xi}\boldsymbol{\xi}') = \boldsymbol{\Phi}$ and $diag(\boldsymbol{\Phi}) = \mathbf{I}_{q \times q}$, $\boldsymbol{\Psi}$ is a diagonal, positive definite loading matrix of unique factors, and $\boldsymbol{\varepsilon}$ is a random vector of unique factors with $E(\boldsymbol{\xi}\boldsymbol{\xi}') = \mathbf{I}_{p \times p}$, $E(\boldsymbol{\xi}\boldsymbol{\varepsilon}') = \mathbf{0}$. Accordingly, the covariance of observed variables is

$$Cov(\mathbf{x}) = E(\mathbf{x}\mathbf{x}') = \boldsymbol{\Sigma} = \mathbf{\Lambda}\boldsymbol{\Phi}\mathbf{\Lambda}' + \boldsymbol{\Psi}. \qquad (2)$$

Thurstone's (1935) regression factor score predictor, often used in software packages is

$$\xi_r = \Phi \Lambda' \Sigma^{-1} \mathbf{x}, \tag{3}$$

has been shown to have maximum determinacy in Krijnen, Wansbeek, and ten Berge (1996) and McDonald and Burr (1967). The inter-correlations of $\xi_r$ are

$$Cor(\xi_r) = diag(\Phi \Lambda' \Sigma^{-1} \Lambda \Phi)^{-1/2} \Phi \Lambda' \Sigma^{-1} \Lambda \Phi \, diag(\Phi \Lambda' \Sigma^{-1} \Lambda \Phi)^{-1/2} \neq \Phi. \tag{4}$$

According to McDonald (1981), the correlation-preserving factor score predictor is

$$\xi_c = \mathbf{N}(\mathbf{N}' \Lambda' \Psi^{-2} \Sigma \Psi^{-2} \Lambda \mathbf{N})^{-1/2} \mathbf{N}' \Lambda' \Psi^{-2} \mathbf{x}, \tag{5}$$

with $\Phi = \mathbf{NN}'$. The inter-correlations of $\xi_c$ are

$$Cor(\xi_c) = \mathbf{N}(\mathbf{N}' \Lambda' \Psi^{-2} \Sigma \Psi^{-2} \Lambda \mathbf{N})^{-1/2} \mathbf{N}' \Lambda' \Psi^{-2} \Sigma \Psi^{-2} \Lambda \mathbf{N}(\mathbf{N}' \Lambda' \Psi^{-2} \Sigma \Psi^{-2} \Lambda \mathbf{N})^{-1/2} \mathbf{N}' = \Phi. \tag{6}$$

From Beauducel and Hilger (2022, Eq. 10) we get

$$\xi_{c2} = \Phi^{1/2} \mathbf{C}_{\xi_s}^{-1/2} diag(\xi_s \xi_s')^{-1/2} \xi_s, \tag{7}$$

with $\xi_s$ as the factor score predictor computed from the software-package at hand and $\mathbf{C}_{\xi_s} = diag(\xi_s \xi_s')^{-1/2} \xi_s \xi_s' diag(\xi_s \xi_s')^{-1/2}$. The inter-correlations of $\xi_s$ are

$$Cor(\xi_{c2}) = \Phi^{1/2} \mathbf{C}_{\xi_s}^{-1/2} \mathbf{C}_{\xi_s} \mathbf{C}_{\xi_s}^{-1/2} \Phi^{1/2} = \Phi. \tag{8}$$

Beauducel and Hilger (2022) consider mean plausible-values from Bayesian confirmatory factor analysis, but any other factor score predictor can be transformed according to Equation 8. From a practical perspective it could be an advantage that the observed scores $\mathbf{x}$ are not needed for the computation of $\xi_{c2}$, whereas they are needed for the computation of $\xi_c$. However, most software-packages compute the regression score predictor, so that $\xi_r$ was entered for $\xi_s$ into Equation 7. According to Beauducel and Hilger (2022), this yields

$$\xi_{c2} = \Phi^{1/2} (\Lambda' \Sigma^{-1} \Lambda)^{-1/2} \Lambda' \Sigma^{-1} \mathbf{x}. \tag{9}$$

The factor score determinacy coefficients can be calculated from post-multiplication of the z-standardized factor score predictor with $\xi$. For $\xi_r$ the determinacy coefficient is

$$\mathbf{P}_{\xi_r} = diag(E(\xi_r \xi')) = diag(\Phi \Lambda' \Sigma^{-1} \Lambda \Phi)^{1/2}. \tag{10}$$

For $\xi_c$ the determinacy coefficient is

$$\mathbf{P}_{\xi_c} = diag(E(\xi_c \xi')) = diag(\mathbf{N}(\mathbf{N}' \Lambda' \Psi^{-2} \Sigma \Psi^{-2} \Lambda \mathbf{N})^{-1/2} \mathbf{N}' \Lambda' \Psi^{-2} \Lambda \Phi), \tag{11}$$

and for $\xi_{c2}$ the determinacy coefficient is

$$\begin{aligned}\mathbf{P}_{\xi_{c2}} &= diag(E(\xi_{c2}\xi')) = diag(\mathbf{\Phi}^{1/2}(\mathbf{\Lambda'}\mathbf{\Sigma}^{-1}\mathbf{\Lambda})^{-1/2}\mathbf{\Lambda'}\mathbf{\Sigma}^{-1}\mathbf{\Lambda}\mathbf{\Phi}) \\ &= diag(\mathbf{\Phi}^{1/2}(\mathbf{\Lambda'}\mathbf{\Sigma}^{-1}\mathbf{\Lambda})^{1/2}\mathbf{\Phi}).\end{aligned} \quad (12)$$

It follows from Equations 4, 6, 8, and 9 that $Cor(\xi_c) = Cor(\xi_{c2}) = \mathbf{\Phi}$ whereas $Cor(\xi_r\xi_r') \neq \mathbf{\Phi}$. It is also known from Krijnen, Wansbeek, and ten Berge (1996) and McDonald and Burr (1967) that $\mathbf{P}_{\xi_r}$ has the largest possible determinacy coefficient. It is therefore of interest to compare the bias of the inter-correlations $Bias(\xi_r) = Cor(\xi_r\xi_r') - \mathbf{\Phi}$ with the loss of determinacy of the correlation-preserving factor score predictors, $Loss(\mathbf{P}_{\xi_c}) = \mathbf{P}_{\xi_c} - \mathbf{P}_{\xi_r}$ and $Loss(\mathbf{P}_{\xi_{c2}}) = \mathbf{P}_{\xi_{c2}} - \mathbf{P}_{\xi_r}$.

## Simulation Study

*Population Simulation*

A simulation study was performed in order to compare the loss of determinacy and the zero-bias in reproducing $\mathbf{\Phi}$ using $\xi_c$ or $\xi_{c2}$ with the gain of determinacy and the bias in reproducing $\mathbf{\Phi}$ using $\xi_r$. The following conditions were investigated in the population: The number of factors $q \in \{3, 6, 9\}$, the size of the salient loadings $sl \in \{.40, .50, .60, .70\}$, the factor inter-correlation $\phi_\xi \in \{.00, .10, .20, .30, .40, .50, .60\}$, and the number of salient loadings per factor $p_{sl} \in \{5, 10\}$. There was one condition without salient loading variability (Var($sl$) = 0) and one condition with non-zero salient loading variability (Var($sl$) > 0). Salient loading variability was introduced by adding or subtracting .05 and .10 to the mean salient loading (see Table 1). There was one condition with zero non-salient loadings ($nl$ = .00) and one condition with non-zero non-salient loadings ($nl$ > .00). Accordingly, there were 3($q$) × 4($sl$) × 7($\phi_\xi$) × 2($p/q$) × 2(Var($sl$)) × 2($nl$) = 672 conditions. Three example loading patterns for mean salient loadings of .50, $p/q$ = 5, and $\phi_\xi$ = .30 are given in Table 1. The pattern of the non-salient loadings was the same across all conditions with non-zero non-salient loadings. The most important dependent variables were $Bias(\xi_r)$, $Loss(\mathbf{P}_{\xi_c})$, and $Loss(\mathbf{P}_{\xi_{c2}})$ although the mean determinacy coefficients and mean inter-correlations are also presented.

Table 1. Example for three conditions based on $sl = .50$ and $\phi_\xi = .30$

|  | $sl = .50$, var($sl$) > 0, $nl$ > 0 | | | $sl = .50$, var($sl$) > 0, $nl$ = 0 | | | $sl = .50$, var($sl$) = 0, $nl$ = 0 | | |
| --- | --- | --- | --- | --- | --- | --- | --- | --- | --- |
|  | $\xi_1$ | $\xi_2$ | $\xi_3$ | $\xi_1$ | $\xi_2$ | $\xi_3$ | $\xi_1$ | $\xi_2$ | $\xi_3$ |
| $x_1$ | **.40** | .10 | -.10 | **.40** | .00 | .00 | **.50** | .00 | .00 |
| $x_2$ | **.45** | .10 | -.10 | **.45** | .00 | .00 | **.50** | .00 | .00 |
| $x_3$ | **.50** | .10 | .10 | **.50** | .00 | .00 | **.50** | .00 | .00 |
| $x_4$ | **.55** | .00 | .10 | **.55** | .00 | .00 | **.50** | .00 | .00 |
| $x_5$ | **.60** | -.10 | .00 | **.60** | .00 | .00 | **.50** | .00 | .00 |
| $x_6$ | -.10 | **.40** | .10 | .00 | **.40** | .00 | .00 | **.50** | .00 |
| $x_7$ | -.10 | **.45** | .10 | .00 | **.45** | .00 | .00 | **.50** | .00 |
| $x_8$ | .10 | **.50** | .10 | .00 | **.50** | .00 | .00 | **.50** | .00 |
| $x_9$ | .10 | **.55** | .00 | .00 | **.55** | .00 | .00 | **.50** | .00 |
| $x_{10}$ | .00 | **.60** | -.10 | .00 | **.60** | .00 | .00 | **.50** | .00 |
| $x_{11}$ | .10 | -.10 | **.40** | .00 | .00 | **.40** | .00 | .00 | **.50** |
| $x_{12}$ | .10 | -.10 | **.45** | .00 | .00 | **.45** | .00 | .00 | **.50** |
| $x_{13}$ | .10 | .10 | **.50** | .00 | .00 | **.50** | .00 | .00 | **.50** |
| $x_{14}$ | .00 | .10 | **.55** | .00 | .00 | **.55** | .00 | .00 | **.50** |
| $x_{15}$ | -.10 | .00 | **.60** | .00 | .00 | **.60** | .00 | .00 | **.50** |
|  | $\phi_\xi$ | | | $\phi_\xi$ | | | $\phi_\xi$ | | |
| $\xi_1$ | 1.00 | | | 1.00 | | | 1.00 | | |
| $\xi_2$ | .30 | 1.00 | | .30 | 1.00 | | .30 | 1.00 | |
| $\xi_3$ | .30 | .30 | 1.00 | .30 | .30 | 1.00 | .30 | .30 | 1.00 |

Note. Loadings ≥ .40 are given in bold-face.

*Results of Population Simulation*

The mean determinacy coefficients and the mean inter-correlations of the factor score predictors across populations based on the same salient loadings are given in Table 2. Obviously, there is only a slight decrease of determinacy for the correlation-preserving factor score predictors $\xi_c$ and $\xi_{c2}$ when compared with $\xi_r$. Both $\xi_c$ and $\xi_{c2}$ have identical determinacy coefficients for each level of *sl*. For small levels of *sl*, there is a substantial over-estimation of the inter-correlations of the factors for $\xi_r$, whereas the mean of the inter-correlations of $\xi_c$ and $\xi_{c2}$ equals the mean of the factor inter-correlations. The effect of the number of factors $q$ was minimal with Mean($\mathbf{P}_{\xi_r}$) = .87 for $q = 3$ and Mean($\mathbf{P}_{\xi_r}$) = .88 for $q = 6$ and 9. Mean($\mathbf{P}_{\xi_c}$) and Mean($\mathbf{P}_{\xi_{c2}}$) were .87 for $q = 3$ and 6 and .88 for $q = 9$.

Table 2. Means of determinacy coefficients and inter-correlations of the factor score predictors (standard deviations in brackets) for the population simulation

| sl | Mean($\mathbf{P}_{\xi_r}$) | Mean($\mathbf{P}_{\xi_c}$) | Mean($\mathbf{P}_{\xi_{c2}}$) | Number of conditions |
|---|---|---|---|---|
| .40 | .80 (.05) | .79 (.05) | .79 (.05) | 168 |
| .50 | .86 (.04) | .86 (.04) | .86 (.04) | 168 |
| .60 | .91 (.03) | .90 (.03) | .90 (.03) | 168 |
| .70 | .94 (.02) | .94 (.02) | .94 (.02) | 168 |
| Total | .88 (.07) | .87 (.07) | .87 (.07) | 672 |
|  | Mean($Cor(\xi_r)$)[a] | Mean($Cor(\xi_c)$)[a] | Mean($Cor(\xi_{c2})$)[a] |  |
| .40 | .41 (.27) | .30 (.20) | .30 (.20) | 168 |
| .50 | .38 (.25) | .30 (.20) | .30 (.20) | 168 |
| .60 | .36 (.24) | .30 (.20) | .30 (.20) | 168 |
| .70 | .34 (.22) | .30 (.20) | .30 (.20) | 168 |
| Total | .37 (.24) | .30 (.20) | .30 (.20) | 672 |

Note. [a]The mean was calculated only for the non-diagonal elements of the correlation matrix.

As the effect of *q* on determinacy coefficients and factor inter-correlations could be neglected and as $\xi_c$ and $\xi_{c2}$ had identical mean determinacy coefficients and inter-correlations, the effects of the remaining conditions were only presented for $q = 3$ and for $Bias(\xi_r)$ and $Loss(\mathbf{P}_{\xi_c})$ (see Figures 1 and 2). Note that the bias for the estimation of the factor inter-correlations is zero for $\xi_c$ and $\xi_{c2}$ whereas the loss of determinacy is zero for $\xi_r$. While $Loss(\mathbf{P}_{\xi_c})$ increased only slightly with increasing $\phi_\xi$, there was a relevant increase of $Bias(\xi_r)$ with increasing $\phi_\xi$. The increase of $Loss(\mathbf{P}_{\xi_c})$ and the increase of $Bias(\xi_r)$ with $\phi_\xi$ was more substantial for smaller levels of *sl*. The effects of Var(*sl*), *nl*, and *p/q* on $Loss(\mathbf{P}_{\xi_c})$ and $Bias(\xi_r)$ were very small. Overall, the bias of the inter-correlations of $\xi_r$ was about twice as large as the loss of determinacy when using $\xi_c$ instead of $\xi_r$.

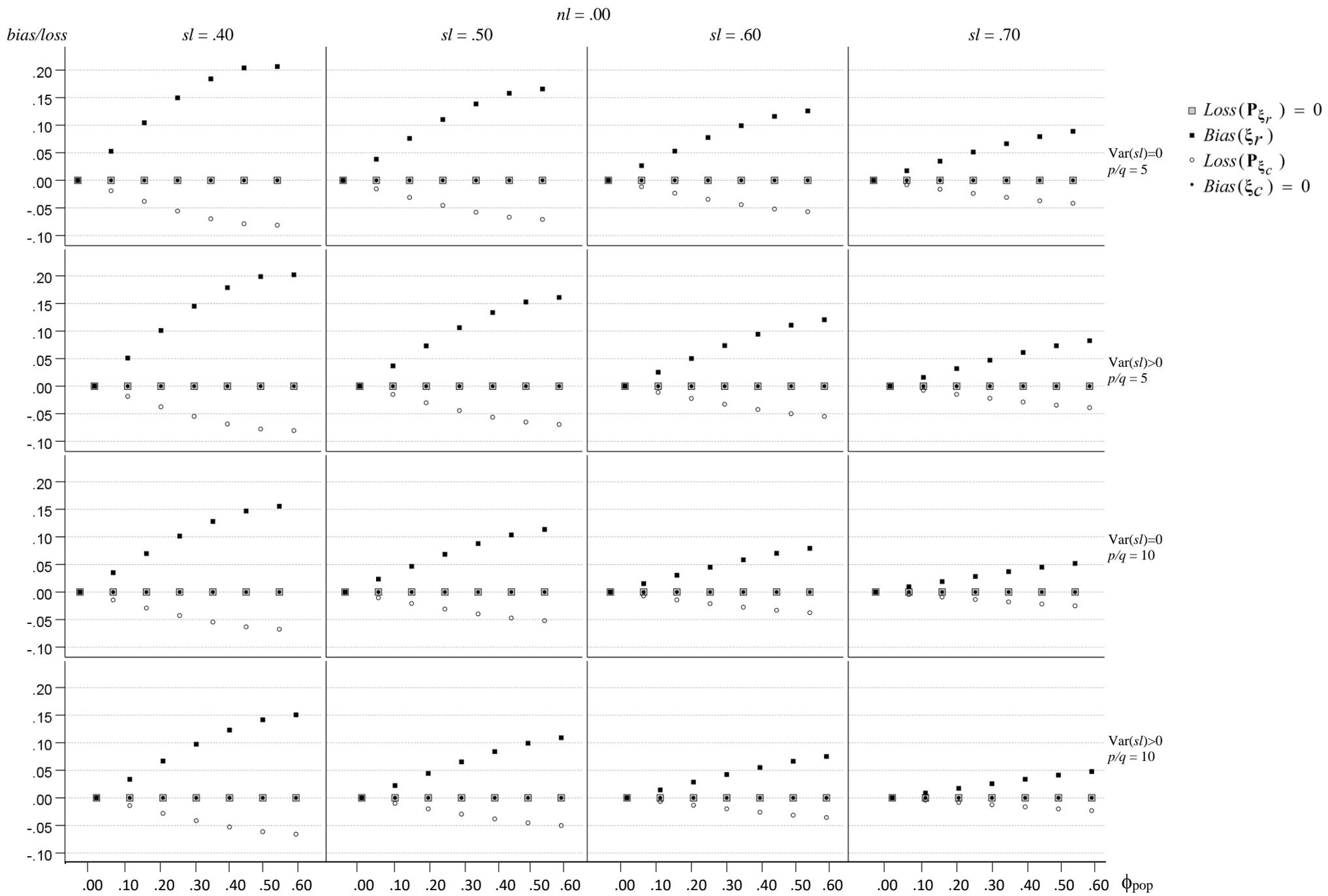

Figure 1. Population simulation of effects of salient loading size (*sl*), salient loading variation (Var(*sl*)), number of salient loadings per factor (*p/q*), zero non-salient loadings (*nl* = .00), and factor inter-correlations ($\phi_{pop}$) on determinacy and inter-correlation of the regression score predictor ($\rho_{reg}, \phi_{reg}$) and of McDonald's correlation-preserving factor score predictor ($\rho_{cor}, \phi_{cor}$).

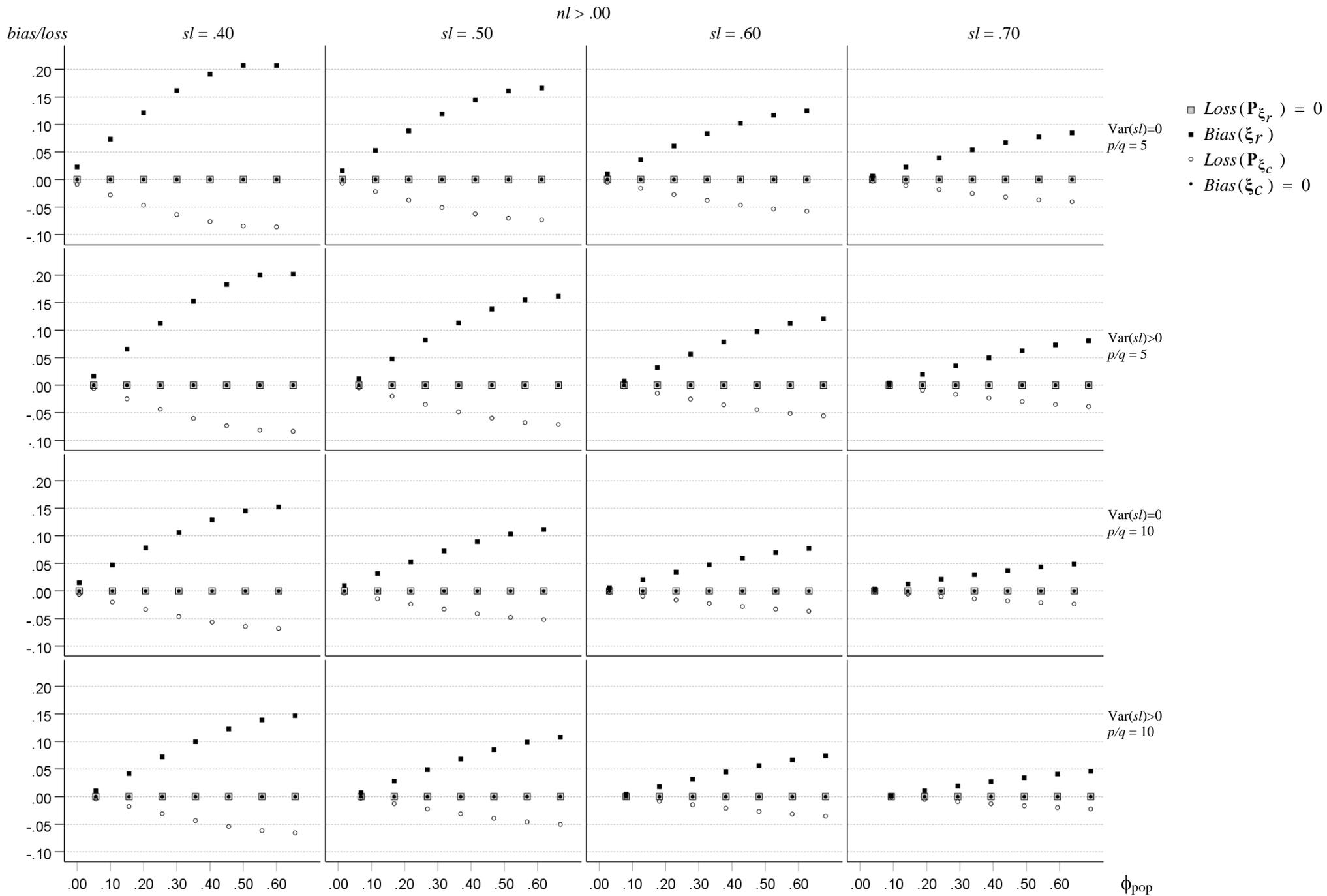

Figure 2. Population simulation of effects of salient loading size (*sl*), salient loading variation (Var(*sl*)), number of salient loadings per factor (*p/q*), non-zero non-salient loadings (*nl* > .00), and factor inter-correlations ($\phi_{pop}$) on determinacy and inter-correlation of the regression score predictor ($\rho_{reg}$, $\phi_{reg}$) and of McDonald's correlation-preserving factor score predictor ($\rho_{cor}$, $\phi_{cor}$).

*Sample Simulation*

In order to investigate the effect of sampling error of the results, the following subset of conditions of the population simulation was investigated in 1,000 samples per condition with $n \in \{300, 600, 900\}$ cases. The number of factors $q \in \{3, 6, 9\}$, the size of the salient loadings $sl \in \{.40, .50, .60\}$, the factor inter-correlation $\phi_\xi \in \{.00, .30, .50\}$ for $p/q = 5$ salient loadings per factor. There was one condition without salient loading variability (Var($sl$) = 0) and one condition with non-zero salient loading variability (Var($sl$) > 0). There was one condition with zero non-salient loadings ($nl$ = .00) and one condition with non-zero non-salient loadings ($nl$ > .00). Salient loading variability and non-zero non-salient loadings were specified as in the population simulation (see Table 1). Accordingly, there were $3(n) \times 3(q) \times 3(sl) \times 3(\phi_\xi) \times 2(\text{Var}(sl)) \times 2(nl) = 324$ conditions in the simulation for the samples.

The common and unique factor scores for each of the 1,000 samples per condition with normal distributions ($\mu = 0$, $\sigma = 1$) were computed by the method of Box and Muller (1958) from uniformly distributed numbers, generated by the Mersenne twister integrated in SPSS. Observed variable scores were computed from the factor scores by means of Equation 1. The observed variables of each sample were submitted to iterative principal axis factor analysis for the extraction of $q$ factors with subsequent oblique Target-rotation (Hurley & Cattell, 1962) towards the population loading pattern. Target-rotation was performed in order to eliminate the effect of different methods of analytic factor rotation on the results.

In order to give an account of the sampling error for the determinacy coefficients and the factor inter-correlations, the dependent variables were the mean and standard deviations of the determinacy coefficients $\mathbf{P}_{\xi_r}$, $\mathbf{P}_{\xi_c}$, and $\mathbf{P}_{\xi_{c2}}$ as well as the mean of the factor score inter-correlations $Cor(\xi_r)$, $Cor(\xi_c)$, and $Cor(\xi_{c2})$.

*Results of Sample Simulation*

Although target-rotation was performed, the mean factor inter-correlations for $n = 300$ and $n = 600$ were substantially smaller than the population factor inter-correlation. As the correlation-preserving factor score predictor was based on the factor inter-correlations estimated from target-rotation, their mean inter-correlation was also smaller than the population factor inter-correlation. However, as the population factor inter-correlation will be unknown in empirical settings, the factor inter-correlations estimated from the factor model were used for the computation of $\xi_c$ and $\xi_{c2}$. Accordingly, differences of $Cor(\xi_r)$ and $Cor(\xi_c)$ or $Cor(\xi_{c2})$ indicate bias of $Cor(\xi_r)$.

The mean determinacy coefficients and the mean inter-correlations of the factor score predictors across samples based on the same salient loadings are given in Table 3. Although mean determinacy (mean $\mathbf{P}_{\xi_r}$) is about .01 larger than the mean $\mathbf{P}_{\xi_c}$ and the mean $\mathbf{P}_{\xi_{c2}}$, the mean inter-correlations (mean $Cor(\xi_r)$) are about .07 larger than the means of $Cor(\xi_c)$ and $Cor(\xi_{c2})$. Thus, nearly identical mean determinacies co-occur with relevant differences regarding the mean inter-correlations.

Table 3. Means of determinacy coefficients and inter-correlations of the factor score predictors (standard deviations in brackets) for the sample simulation

| Sl | Mean($\mathbf{P}_{\xi_r}$) | Mean($\mathbf{P}_{\xi_c}$) | Mean($\mathbf{P}_{\xi_{c2}}$) | Number of conditions |
|---|---|---|---|---|
| .40 | .69 (.10) | .68 (.10) | .68 (.10) | 108 |
| .50 | .81 (.03) | .80 (.03) | .80 (.03) | 108 |
| .60 | .88 (.01) | .87 (.01) | .87 (.01) | 108 |
| Total | .79 (.10) | .79 (.10) | .79 (.10) | 324 |
|  |  |  |  | 324 |
|  | Mean($Cor(\xi_r)$)[a] | Mean($Cor(\xi_c)$)[a] | Mean($Cor(\xi_{c2})$)[a] |  |
| .40 | .27 (.23) | .18 (.16) | .18 (.16) | 108 |
| .50 | .31 (.24) | .23 (.18) | .23 (.18) | 108 |
| .60 | .31 (.24) | .25 (.19) | .25 (.19) | 108 |
| Total | .30 (.24) | .22 (.18) | .22 (.18) | 324 |

Note. [a]The mean was calculated only for the non-diagonal elements of the correlation matrix.

More detailed results are presented in Figure 3 and 4. However, as the results for $\xi_c$ and $\xi_{c2}$ were identical, only the results for $\xi_c$ were presented in the figures. For the conditions with $\phi_{pop}$ = .00 and $nl$ = .00, the means of $\mathbf{P}_{\xi_r}$ and $\mathbf{P}_{\xi_c}$ are nearly identical and the means of $Cor(\xi_r)$ and $Cor(\xi_c)$ are close to zero. Thus, there is no relevant difference between $\xi_r$ and $\xi_c$ for these conditions (see Figure 3).

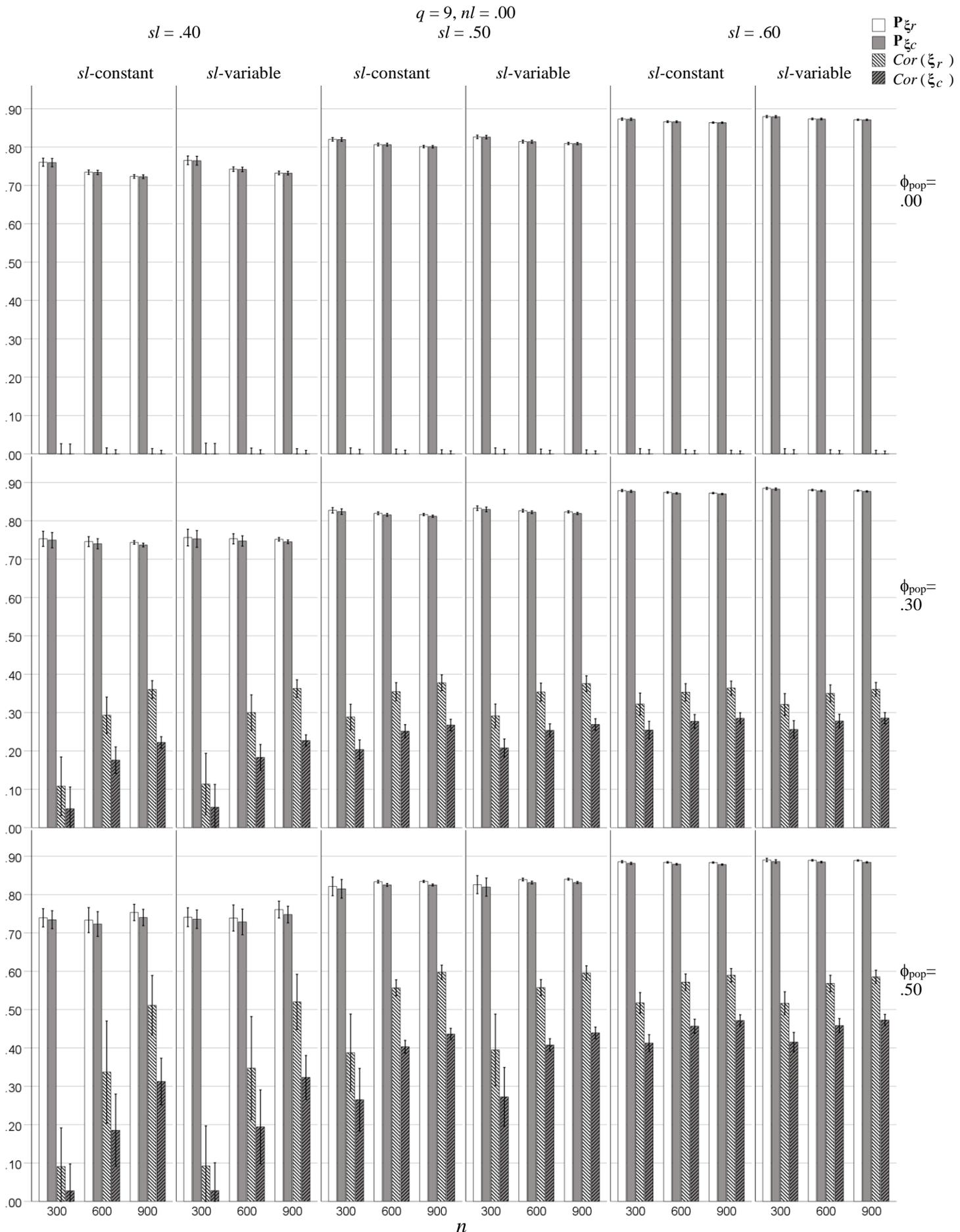

Figure 3. Mean($\mathbf{P}_{\xi_r}$), Mean($\mathbf{P}_{\xi_c}$), and mean inter-correlation of regression factor score predictor, $Cor(\xi_r)$, and of correlation-preserving factor score predictor, $Cor(\xi_c)$, for $q = 9$ factors and zero non-salient loadings ($nl = .00$); $sl$ = salient loadings; $\phi_{pop}$ = population factor correlation; the error bars show the standard deviation.

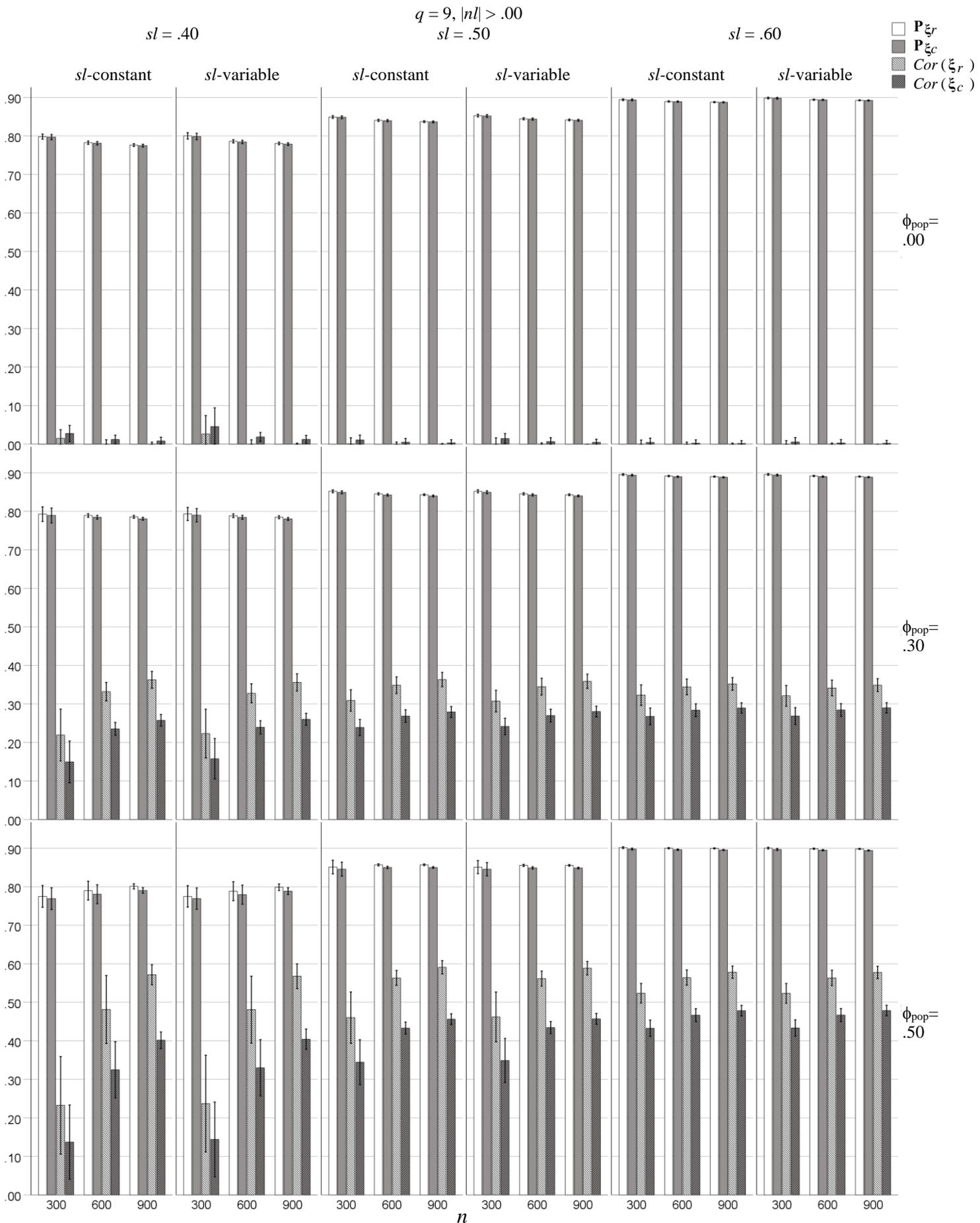

Figure 4. Mean($\mathbf{P}_{\xi_r}$), Mean($\mathbf{P}_{\xi_c}$), and mean inter-correlation of regression factor score predictor, $Cor(\xi_r)$, and of correlation-preserving factor score predictor, $Cor(\xi_c)$, for $q = 9$ factors and substantial non-salient loadings ($|nl| > .00$); $sl$ = salient loadings; $\phi_{pop}$= population factor correlation; the error bars show the standard deviation.

14For larger $\phi_{pop}$ the overestimation of the inter-correlations of the factors by means of $Cor(\xi_r)$ becomes more pronounced whereas the loss of determinacy in $\mathbf{P}_{\xi_c}$ remains rather small. The effects are similar when non-salient population loadings are zero (Figure 3) or when small non-zero non-salient population loadings occur (Figure 4).

## Empirical example

A subsample of 242 participants (126 female, aged $M = 8.6$ and $SD = .68$ years) without missing values in three measurement occasions in intervals of six months was drawn from a large study on elementary school children. The main goal of this study was to find predictors of writing performance (Kuhl, 2020). In this context gender stereotypes about family roles were assessed. At each measurement occasion, five Likert-type items measure the participants' attitude towards these stereotypes (Hoffman & Kloska, 1995; Meurer, 2012). Agreement of parents and schools was obtained by declaration of consent after receiving an information letter. Data collection was conducted at school.

*Results*

A confirmatory factor model with a gender stereotype factor for each of the three measurement occasions and four correlated errors was estimated by means of maximum likelihood estimation with Mplus 8.4 (Muthén & Muthén, 2019). The model fits well to the data and the factor inter-correlations are rather high so that it could be an issue that the inter-correlations of the regression factor score predictor are substantially larger than the factor inter-correlations (see Figure 5).

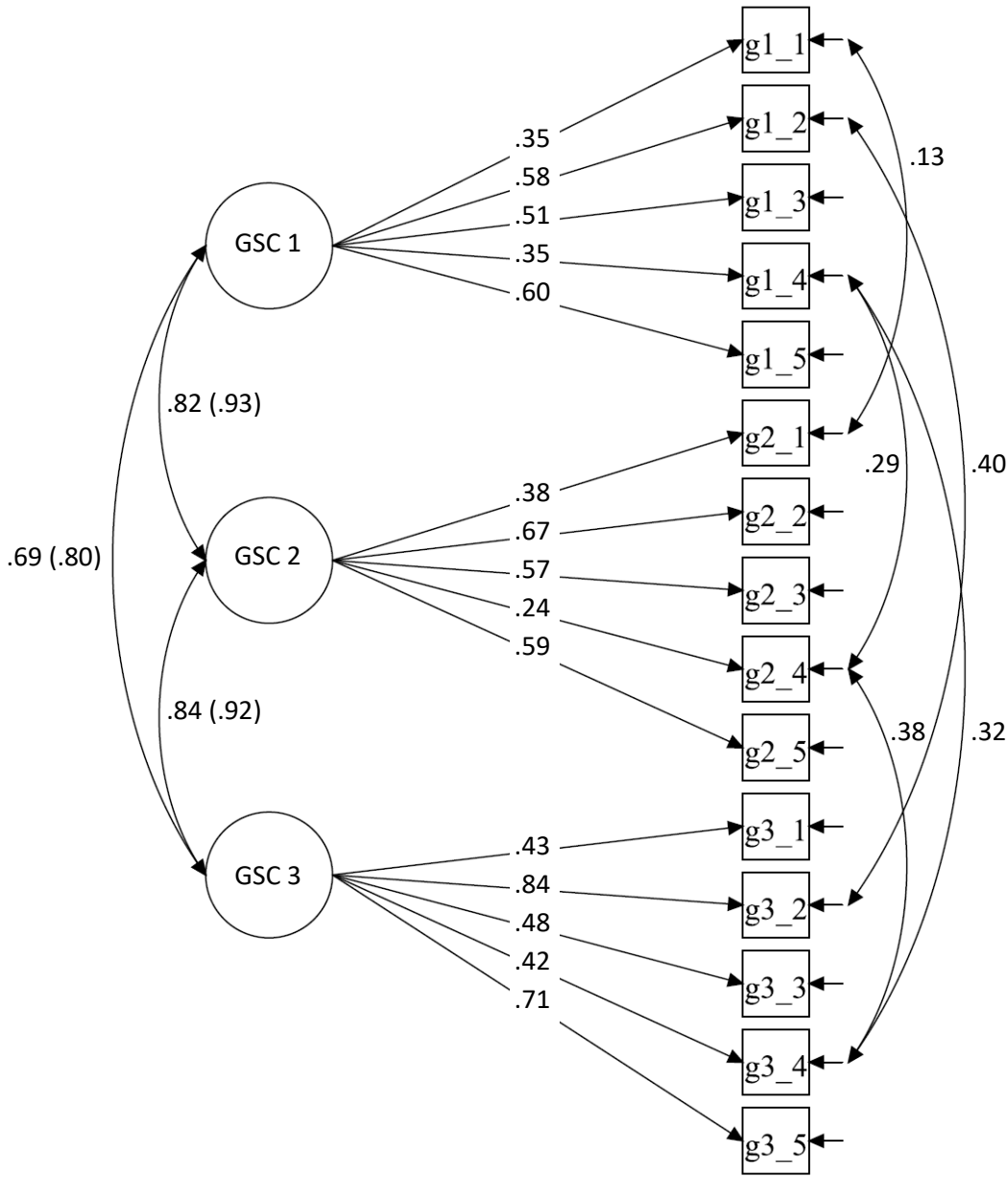

Figure 5. Confirmatory factor model (completely standardized solution) for gender stereotype concept (GSC) at three measurement occasions, model fit: $\chi^2(82)= 125.81$, $p < .01$; CFI = .93, RMSEA = .047, SRMR = .054; the inter-correlations of the regression factor scores computed by Mplus 8.4 are given in brackets behind the factor inter-correlations.

For each factor, the determinacy coefficient of the regression factor score predictor is only slightly larger than the corresponding determinacy coefficient of the correlation-preserving factor score predictors (see Table 4). The SPSS-script for the computation of $\xi_{c2}$ from $\xi_r$ as well as the determinacies and inter-correlations of $\xi_r$ and $\xi_{c2}$ is given in the Appendix. Since the inter-correlations of the correlation-preserving factor scores are identical to the factor inter-correlations, they needed not to be presented separately. To sum up, on average the inter-correlations of the regression factor score predictor are about .10 larger than the inter-correlations of the correlation-preserving factor score predictors whereas the determinacy coefficients of the regression factor score predictor are on average about .01 larger than the determinacy coefficients of the correlation-preserving factor score predictors for the corresponding factors.

Table 4. Empirical example: Determinacy coefficients for regression factor scores ($\mathbf{P}_{\xi_r}$) and for correlation-preserving factor scores ($\mathbf{P}_{\xi_c}$, $\mathbf{P}_{\xi_{c2}}$)

| Factor | $\mathbf{P}_{\xi_r}$ | $\mathbf{P}_{\xi_c}$ | $\mathbf{P}_{\xi_{c2}}$ |
|---|---|---|---|
| GSC 1 | .85 | .84 | .84 |
| GSC 2 | .89 | .88 | .88 |
| GSC 3 | .92 | .91 | .92 |

## Discussion

The aim of the present study was to compare the regression score predictor $\xi_r$ with the correlation-preserving factor score predictors $\xi_c$ and $\xi_{c2}$ regarding the determinacy coefficients and the inter-correlations. A population simulation study for 324 population model parameters and a simulation study for samples based on 108 population model parameters indicate that the determinacy coefficients of $\xi_c$ and $\xi_{c2}$ are slightly smaller than the determinacy coefficients of $\xi_r$, while the overestimation of inter-correlations is substantial for $\xi_r$ while it does not occur for $\xi_c$ and $\xi_{c2}$.

The empirical example shows that the gender stereotypes can be measured by the five items. The inter-correlations of error terms only refer to correlations of identical items at different measurement occasions so that they probably represent item-specific content. The inter-correlations of $\xi_r$ can be substantially larger than the factor inter-correlations, especially, when the factor inter-correlations are already large. Thereby, the use of $\xi_r$ may impair discriminant validity. Moreover, the determinacy coefficients of $\xi_r$ were only slightly larger than the determinacy coefficients of $\xi_c$ and $\xi_{c2}$ for the corresponding factors. The empirical example therefore shows that slightly larger determinacy coefficients for $\xi_r$ than for $\xi_c$ and $\xi_{c2}$ can go along with substantial bias of the inter-correlations for $\xi_r$ while the inter-correlations of $\xi_c$ and $\xi_{c2}$ are unbiased.

Although the results of the simulation studies and the empirical example indicate that $\xi_c$ and $\xi_{c2}$ might be preferred over $\xi_r$ in several settings, the decision, which aspect is more important depends on the intended use of the factor score predictors. When an immediate decision is based on the scores, one might nevertheless prefer the scores with the highest determinacy, i.e., $\xi_r$. However, when the scores are entered as predictors in further investigations, the more precise representation of the inter-correlations of the factors might be more important, so that $\xi_c$ and $\xi_{c2}$ might be preferred. As the results for $\xi_c$ and $\xi_{c2}$ were nearly identical, it might depend on pragmatic reasons which one of these scores is preferred. The advantage of $\xi_{c2}$ is that it can be computed directly from $\xi_r$ so that it is not necessary to have the observed variables. Overall, the decision on the factor score predictor that should be preferred should take the possible trade of between the maximum determinacy and the maximum precision of reproducing the factor inter-correlations into account. As a basis for a decision, the determinacies of $\xi_r$, and $\xi_c$ or $\xi_{c2}$ should be compared as should their respective inter-correlations. An SPSS script is given in the Appendix in order to compute $\xi_{c2}$ from $\xi_r$ and the determinacy and the inter-correlations of $\xi_r$ and $\xi_{c2}$ so that it is possible to decide on a rational basis which factor score predictor should be used. The results of the present study indicate that such an informed decision will be an improvement over the standard use of $\xi_r$ which is typically the default of software packages.

# References

Asparouhov, T., & Muthén B. (2010). *Plausible values for latent variables using Mplus.* Technical Report. Retrieved from www.statmodel.com/download/Plausible.pdf

Beauducel, A. & Hilger, N. (2022). Correlation-preserving mean plausible values as a basis for prediction in the context of Bayesian structural equation modeling. *International Journal of Statistics and Probability; 11(6)*, 1-11. https://doi.org/10.5539/ijsp.v11n6p1

Beauducel, A., & Hilger, N. (in press). Coefficients of factor score determinacy for mean plausible values of Bayesian factor analysis. *Educational and Psychological Measurement*, 2022. https://doi.org/10.1177/00131644221078960

Box, G. E. P., & Muller, M. E. (1958). A note on the generation of random normal deviates. *Annals of Mathematical Statistics, 29(2)*, 610-611. https://doi.org/10.1214/aoms/1177706645

DiStefano, C., Zhu, M., & Mîndrilă, D. (2009). Understanding and using factor scores: Considerations for the applied researcher. *Practical Assessment, Research & Evaluation, 14*, 1-11. Available online: http://pareonline.net/getvn.asp?v=14&n=20

Grice, J. W. (2001). Computing and evaluating factor scores. *Psychological Methods, 6,* 430-450. https://doi.org/10.1037/1082-989X.6.4.430

Hoffman, L. W. & Kloska, D. D. (1995). Parents' gender-based attitudes toward marital roles and child rearing: Development and validation of new measures. *Sex Roles, 32(5)*, 273-295. https://doi.org/10.1007/BF01544598

Hurley, J. R. & Cattell, R. B. (1962). The Procrustes program: Producing direct rotation to test a hypothesized factor structure. *Behavioral Science, 7(2)*, 258. https://doi.org/10.1002/bs.3830070216

Krijnen, W. P., Wansbeek, T., & ten Berge, J.M.F. (1996). Best linear predictors for factor scores. *Communications in Statistics - Theory and Methods, 25,* 3013–3025. https://doi.org/10.1080/03610929608831883

Kuhl, T. (2020). *Rechtschreibung in der Grundschule. Eine empirische Untersuchung der Auswirkungen verschiedener Unterrichtsmethoden* [Orthography in elementary school. An empirical evaluation of different teaching methods]. Wiesbaden: Springer. https://doi.org/10.1007/978-3-658-29908-8

McDonald, R. P. (1981). Constrained least squares estimators of oblique common factors. *Psychometrika, 46*, 337-341. https://doi.org/10.1007/BF02293740

McDonald, R. P. & Burr, E.J. (1967). A comparison of four methods of constructing factor scores. *Psychometrika, 32*, 381-401. https://doi.org/10.1007/BF02289653

Meurer, J. (2012). *Schuldgefühle bei berufstätigen Müttern. Eine qualitative Kinderbefragung* [Feelings of guilt among working mothers with preschool children] (Diplomarbeit, University of Bonn).

Nicewander, W. A. (2020). A Perspective on the Mathematical and Psychometric Aspects of Factor Indeterminacy. *Multivariate Behavioral Research, 55(6)*, 825-838. https://doi.org/10.1080/00273171.2019.1684872

Sörbom, D. (1974). A general method for studying differences in factor means and factor structure between groups. *British Journal of Mathematical and Statistical Psychology, 27*, 229–239.

Thurstone, L. L. (1935). *The Vectors of mind.* Chicago, IL: University of Chicago Press.

**SPSS-Skript for transformation of regression factor score predictor into correlation-preserving factor score predictor.**

```
* Encoding: windows-1252.

MATRIX.

* Enter the regression factor score predictor here.
get Ksi_r /variables= x1 x2 x3 /file='#####1.sav'.
compute Ksi_r=t(Ksi_r).

* Enter the factor loading matrix here.
compute L={
     0.346,  0.000,  0.000;
     0.582,  0.000,  0.000;
     0.512,  0.000,  0.000;
     0.347,  0.000,  0.000;
     0.598,  0.000,  0.000;
     0.000,  0.377,  0.000;
     0.000,  0.674,  0.000;
     0.000,  0.565,  0.000;
     0.000,  0.245,  0.000;
     0.000,  0.588,  0.000;
     0.000,  0.000,  0.434;
     0.000,  0.000,  0.841;
     0.000,  0.000,  0.480;
     0.000,  0.000,  0.422;
     0.000,  0.000,  0.713
     }.

* Enter the factor inter-correlations here.
compute Phi={
     1.000,  0.822,  0.686;
     0.822,  1.000,  0.838;
     0.686,  0.838,  1.000
     }.

compute p=nrow(L).
compute q=ncol(L).
compute Psi=Mdiag(Diag(1-L*Phi*T(L)))&**0.5.
```

```
* If there are correlated errors, enter the "Model Estimated Correlations" in the lower triangle.
compute Sig={
    1.000,  0.000,  0.000,  0.000,  0.000,  0.000,  0.000,  0.000,  0.000,  0.000,  0.000,  0.000,  0.000,  0.000,  0.000;
    0.201,  1.000,  0.000,  0.000,  0.000,  0.000,  0.000,  0.000,  0.000,  0.000,  0.000,  0.000,  0.000,  0.000,  0.000;
    0.177,  0.298,  1.000,  0.000,  0.000,  0.000,  0.000,  0.000,  0.000,  0.000,  0.000,  0.000,  0.000,  0.000,  0.000;
    0.120,  0.202,  0.178,  1.000,  0.000,  0.000,  0.000,  0.000,  0.000,  0.000,  0.000,  0.000,  0.000,  0.000,  0.000;
    0.207,  0.348,  0.307,  0.207,  1.000,  0.000,  0.000,  0.000,  0.000,  0.000,  0.000,  0.000,  0.000,  0.000,  0.000;
    0.222,  0.181,  0.159,  0.108,  0.186,  1.000,  0.000,  0.000,  0.000,  0.000,  0.000,  0.000,  0.000,  0.000,  0.000;
    0.192,  0.322,  0.284,  0.192,  0.331,  0.254,  1.000,  0.000,  0.000,  0.000,  0.000,  0.000,  0.000,  0.000,  0.000;
    0.161,  0.271,  0.238,  0.161,  0.278,  0.213,  0.381,  1.000,  0.000,  0.000,  0.000,  0.000,  0.000,  0.000,  0.000;
    0.070,  0.117,  0.103,  0.334,  0.121,  0.093,  0.165,  0.139,  1.000,  0.000,  0.000,  0.000,  0.000,  0.000,  0.000;
    0.167,  0.281,  0.248,  0.168,  0.289,  0.222,  0.396,  0.332,  0.144,  1.000,  0.000,  0.000,  0.000,  0.000,  0.000;
    0.103,  0.173,  0.153,  0.103,  0.178,  0.137,  0.245,  0.206,  0.089,  0.214,  1.000,  0.000,  0.000,  0.000,  0.000;
    0.199,  0.510,  0.295,  0.200,  0.345,  0.266,  0.475,  0.398,  0.173,  0.414,  0.365,  1.000,  0.000,  0.000,  0.000;
    0.114,  0.192,  0.169,  0.114,  0.197,  0.152,  0.271,  0.227,  0.099,  0.237,  0.208,  0.404,  1.000,  0.000,  0.000;
    0.100,  0.168,  0.148,  0.369,  0.173,  0.133,  0.238,  0.200,  0.418,  0.208,  0.183,  0.355,  0.203,  1.000,  0.000;
    0.169,  0.284,  0.251,  0.170,  0.292,  0.225,  0.403,  0.338,  0.147,  0.351,  0.310,  0.599,  0.342,  0.301,  1.000
    }.
compute Sig=Sig+t(Sig)-ident(nrow(L),nrow(L)).

* If there are no correlated errors, the "Model Estimated Correlations" can directly be computed from the model parameters (next line).
*compute Sig=L*Phi*T(L)+Psi&**2.

* Transform regression scores into Ksi_c2.
compute D=Mdiag(diag(Ksi_r*t(Ksi_r))&/(ncol(Ksi_r)-1))&**0.5.
compute Ksi_r=INV(D)*Ksi_r.
compute CKsi_r=Ksi_r*t(Ksi_r)&/(ncol(Ksi_r)-1).
CALL SVD(CKsi_r,UU,S,VV).
compute CKsi_r12=UU*(S&**0.5)*t(VV).
compute oKsi_r=INV(CKsi_r12)*Ksi_r.
CALL SVD(Phi,UU,S,VV).
compute Phi12=UU*(S&**0.5)*t(VV).
compute Ksi_c2=Phi12*oKsi_r.
compute CKsi_c2=Ksi_c2*t(Ksi_c2)&/(ncol(Ksi_c2)-1).

* Compute determinacies.
compute PKsi_r=Mdiag(diag(Phi*t(L)*INV(Sig)*L*Phi))&**0.5.
compute LiSL=t(L)*INV(Sig)*L.
CALL SVD(LiSL,UU,S,VV).
compute LiSL12=UU*(S&**0.5)*t(VV).
compute PKsi_c2=Mdiag(diag(Phi12*LiSL12*Phi)).
```

```
* Print and save results.
compute Lab={make(1,ncol(PKsi_r),{"PKsi_r"}),make(1,ncol(PKsi_c2),{"PKsi_c2"})}.
print {PKsi_r,PKsi_c2} /Title="Determinacies:" /format=F8.3 /Cnames=Lab.
compute Lab={make(1,ncol(CKsi_r),{"CKsi_r"}),make(1,ncol(CKsi_c2),{"CKsi_c2"})}.
print {CKsi_r,CKsi_c2} /Title="Inter-correlations:" /format=F8.3 /Cnames=Lab.
save {t(Ksi_c2)} /OUTFILE="#####2.sav".

END MATRIX.

GET FILE="#####2.sav".
```